\documentstyle[12pt]{article}
\renewcommand \baselinestretch{1.2}

\begin{document}

\hfill CEAB 95/4-10


\vspace*{3mm}

\begin{center}

{\LARGE \bf
General dilatonic gravity with an asymptotically free gravitational
coupling constant near two dimensions}

\vspace{8mm}

\renewcommand
\baselinestretch{0.5}
\medskip

{\sc E. Elizalde}
\footnote{E-mail: eli@zeta.ecm.ub.es} \\
Center for Advanced Studies CEAB, CSIC, Cam\'{\i} de Santa
B\`arbara,
17300 Blanes,
\\ and Department ECM and IFAE, Faculty of Physics,
University of Barcelona, \\ Diagonal 647, 08028 Barcelona,
Catalonia, Spain \\
 and \\
{\sc S.D. Odintsov} \footnote{E-mail: odintsov@ecm.ub.es.
On
leave from: Tomsk Pedagogical Institute, 634041 Tomsk, Russia.}
\\
Department ECM, Faculty of Physics,
University of  Barcelona, \\  Diagonal 647, 08028 Barcelona,
Catalonia,
Spain \\

\vspace{15mm}

{\bf Abstract}

\end{center}

We study a renormalizable, general theory of dilatonic gravity (with a
kinetic-like term for the dilaton) interacting with scalar matter near
two dimensions. The one-loop effective action and the beta functions for
this general theory are written down. It is proven that the theory
possesses a
non-trivial ultraviolet fixed point which yields an asymptotically free
gravitational
coupling constant (at $\epsilon \rightarrow 0$) in this regime.
Moreover, at the fixed point the theory can be cast
under the form of a string-inspired model with free scalar matter. The
renormalization of the Jackiw-Teitelboim model and of lineal
gravity in $2+\epsilon$ dimensions is also discussed. We show that these
two theories are distinguished at the quantum level.
Finally, fermion-dilatonic gravity near two dimensions is considered.

\vspace{4mm}


\newpage

\section{  Introduction.}
As is well known, quantum gravity (QG) based on the four-dimensional
Einstein action is not renormalizable \cite{18} (for a detailed account
of the operator formalism in Einstenian gravity, see \cite{27}). Among
the different proposals for the construction of a consistent theory of
QG, the $2+\epsilon$ approach offers a quite interesting way to
solve this problem of nonrenormalizability of Einstein gravity.
In fact, it was shown some time ago \cite{17} that the gravitational
coupling
constant $G$ in $2+\epsilon$ dimensions exhibits an asymptotically free
behavior. But, unfortunately, it has been also shown that
Einstein gravity in
$2+\epsilon$ dimensions (a theory that has no smooth $\epsilon
\rightarrow 0$
limit) is not renormalizable near two dimensions \cite{19}. This renders
the issue of asymptotic freedom of Einstein's theory near two dimensions
rather hopeless.

On the other hand, there has been recently a lot of activity about
two-dimensional quantum dilatonic gravity (see, for example,
\cite{11}-\cite{6}, and for  reviews and more complete lists of
references \cite{13,14}). Dilatonic gravities appear mainly as
string-inspired models and it is common belief that they can provide
very nice toy models for a realistic theory of QG and a reliable
description of two-dimensional black holes.

Dilatonic gravity of a general form is a renormalizable theory
\cite{6,12}. Unlike Einstein gravity, it has a smooth
 $\epsilon \rightarrow 0$ limit. As it was shown recently \cite{4}
---using as example dilatonic gravity without a kinetic term for the
dilaton and without dilatonic potential--- there exists a regime of the
theory where the gravitational
coupling constant is asymptotically free and the dilatonic coupling
functions possess a non-trivial ultraviolet fixed point. As was seen
in Ref. \cite{4}, at the fixed point the theory can be transformed to
the CGHS form \cite{11}.

In the present paper we study the general theory of dilatonic gravity
---which includes in the classical action a kinetic-like term for the
dilaton and a dilatonic potential--- near two dimensions. In the next
section we start from
the classical action and obtain the one-loop effective action in the
covariant gauge. The renormalization group $\beta$-functions for the
gravitational constant $G$ and for the dilatonic couplings are
constructed
to first order in $G$. With the linear {\it Ansatz} for the dilatonic
couplings and in the regime of asymptotic freedom for $G$ the fixed
points of the RG equations are found. It is shown that, at the
non-trivial ultraviolet fixed point, the theory can be presented as
string-inspired dilatonic gravity (not in the CGHS form)
with free scalars. In Sect. 3 a similar study is carried out for
fermion-dilatonic gravity.
In section 4 the one-loop renormalization of two particular models
---namely
the Jackiw-Teitelboim one and lineal gravity--- are discussed near two
dimensions. It is shown that, in a Landau type gauge, lineal gravity is
one-loop renormalizable but only in exactly two dimensions while, on
the contrary, the Jackiw-Teitelboim model is not one-loop
renormalizable, neither in exactly two nor near two dimensions.
\bigskip

\section{General dilatonic gravity in $2+\epsilon$
dimensions.}
We shall consider the theory of dilatonic gravity characterized by the
following Lagrangian \cite{1,12}
 \begin{equation}
L = \frac{\mu^\epsilon}{2} e^{-2Z(\phi)} g^{\mu\nu}
\partial_\mu \phi\partial_\nu \phi + {\mu^\epsilon \over 16\pi
G} e^{-2\phi} R+ \mu^\epsilon m^2 e^{-V(\phi)}
- {1 \over 2} e^{-2 \Phi (\phi)} g^{\mu\nu} \partial_\mu \chi_i
\partial_\nu \chi^i,
 \label{1}
\end{equation}
where $\mu$ is a mass parameter that keeps the correct dimensions in
($2+\epsilon$)-dimensional space, $G$ is the gravitational
coupling, $g_{\mu\nu}$  the ($2+\epsilon$)-dimensional
metric, $R$ the corresponding curvature, $\chi_i$ are scalars ($i=1,2,
\ldots, n$), and where the dilatonic functions $Z(\phi)$,
$V(\phi)$ and $\Phi (\phi)$ are supposed to be smooth enough and chosen
so that $Z(0)=V(0)=\Phi (0)=0$. The mass term $m^2$ is introduced
in (\ref{1}) in order to keep $V(\phi)$
 dimensionless.
 Notice that an arbitrary dilatonic coupling function in front of $R$ in
(\ref{1}) can be always writen (without loss of generality) under the
form chosen here, by a simple redefinition of the dilaton.
Our aim in this paper will be to study in detail the renormalization
structure of (\ref{1}) and, in particular, its renormalization group
near two dimensions.
A very interesting observation is the fact that, unlike
($2+\epsilon$)-dimensional Einstein theory, the theory (\ref{1}) is
renormalizable.
Moreover, it has a smooth limit for $\epsilon \rightarrow
0$. This property allows for the possibility to study the behavior
of (\ref{1}) in $2+\epsilon$ dimensions by simply using the
counterterms  calculated already in two dimensions ---in close
analogy with quantum field theory in frames of the
$\epsilon$-expansion technique (for a review see \cite{2}
and \cite{3}). It is interesting to note that, in the limit $\epsilon
\rightarrow 0$, the theory \cite{1} can be discussed in a string
effective action manner \cite{15,16}.

Studies of dilatonic gravity near two dimensions have been also carried
out in Refs. \cite{4,5}. However, we should point out that in these
works the first term in (\ref{1}) has been chosen to be zero. The
motivation for such choice was the fact that by a conformal
transformation of the metric in (\ref{1}) one can eliminate the kinetic
term for the dilaton and hence the two theories appear to be equivalent.
However, this is only a classical equivalence, that can be easily lost
at the quantum level, as is well known. In consequence, we prefer to
discuss the Lagrangian (\ref{1}) that yields the most general action for
a renormalizable theory of dilatonic gravity.

We choose the covariant gauge-fixing action in the form \cite{6}
\begin{equation}
S_{gf} = -  {\mu^\epsilon \over 32\pi G} \int d^d x\,
\sqrt{-g}\,
 g_{\mu\nu} \chi^\mu \chi^\nu  e^{-2\phi}, \label{2}
\end{equation}
where
\begin{equation}
\chi^\mu =   \nabla_\nu \bar{h}^{\mu\nu} + 2 \nabla^\mu \varphi
\label{2a}
\end{equation}
and $\bar{h}^{\mu\nu}$ is a traceless quantum gravitational field
and $\varphi$ a quantum scalar field, in the background field
method (see \cite{10} for an introduction).

In this gauge (\ref{2}),
the calculation of the one-loop effective action can be done in exactly
two dimensions, with the following result (we shall drop the details of
the evaluation, since these techniques are quite well known by now, see
\cite{1,6,7,7a})
\begin{eqnarray}
\Gamma_{div} &=& \frac{1}{4\pi \epsilon}  \int d^d x\,
\sqrt{-g}\,
\left\{ \frac{24-n}{6} R +16 \pi G m^2 e^{2\phi - V(\phi)} [2 +
V'(\phi)] \right. \nonumber \\
&& \left.  - \left[ 8-n \Phi'(\phi)^2
 +16 \pi G  e^{2\phi - 2Z(\phi)} (2 - Z'(\phi)) \right]
g^{\mu\nu} \partial_\mu \phi \partial_\nu \phi \right\},
\label{3}
\end{eqnarray}
where $\epsilon =d-2$. Thus we have got the one-loop effective action,
in the gauge (\ref{2}), in $2+\epsilon$ dimensions.
For $e^{-2Z(\phi)} =0$ and $\Phi (\phi)=$ const. the one-loop effective
action (\ref{3}) coincides with the one obtained in Refs. \cite{6,7} in
the same gauge, and for the action (\ref{1}) it coincides with the
result obtained in Ref. \cite{1}. The gauge dependence of (\ref{3}) may
be studied in a way similar to that of the last Ref. \cite{6}.

The corresponding counterterms can be written as
\begin{equation}
\Gamma_{count} =- \mu^\epsilon  \int d^d x\,  \sqrt{-g}\,
\left[
RA_1 + g^{\mu\nu} \partial_\mu \phi \partial_\nu \phi
\bar{A}_2(\phi) + m^2A_3(\phi) \right],
\label{4}
\end{equation}
where
\begin{eqnarray}
A_1 &=& \frac{24-n}{24\pi\epsilon}, \nonumber \\
\bar{A}_2(\phi) &=& \frac{1}{4\pi\epsilon} \left[ -8+n\Phi'(\phi)^2
-16 \pi G  e^{2\phi - 2Z(\phi)} (2 - Z'(\phi)) \right] \equiv A_2 (\phi)
+ e^{-2Z(\phi )} A(\phi ), \nonumber \\
A_3(\phi) &=& \frac{4G}{\epsilon}[2 + V'(\phi)]e^{2\phi -V(\phi)}
\equiv e^{-V(\phi)} \widetilde{A}_3 (\phi),
\label{5}
\end{eqnarray}
where the term with $Z(\phi )$ is specified separately.

We can now start with the study of renormalization.
It can be done in close analogy with the approach developed in Refs.
\cite{4,5}, so no details will be given.
Observe that the renormalization of the kinetic term for the dilaton,
i.e. $A_2 (\phi )$, originated from the renormalization of the metric,
as in Refs. \cite{4,5}, and from the contribution from $Z(\phi )$ itself
(which is of the order of $G$). Hence, the only natural way to
renormalize the theory is to consider the renormalization of $Z$ in a
homogeneous way (via $Z$ itself). Then, in the absence of $Z$ in
(\ref{1}) from the very beginning we are back to the situation of Refs.
\cite{4,5}. With these remarks the beta-functions for the dilatonic
coupling functions can be easily obtained.

A careful calculation of the dilatonic beta-functions gives the
following result (see also \cite{4,5} for a discussion of the RG in the
absence of the dilatonic kinetic term in (\ref{1}), i.e. when $e^{-2Z}
=0$)
\begin{eqnarray}
\beta_G &=& \mu {\partial G \over \partial \mu} = \epsilon G - 16
\pi \epsilon A_1 G^2, \label{6} \\
\beta_\Phi (\phi_0) &=& \mu {\partial \Phi (\phi_0)  \over
\partial \mu} = 8\pi \epsilon A_1 G (e^{2\phi_0}-1)\Phi'
(\phi_0) \nonumber \\
&&+ \frac{2\pi \epsilon^2 G}{\epsilon +1} [\Phi' (\phi_0) -1]
\int_0^{\phi_0} d\phi' \, e^{2\phi'}A_2(\phi'),  \label{7} \\
\beta_V (\phi_0) &=& \mu {\partial V (\phi_0)  \over \partial
\mu}
= 8\pi \epsilon A_1 G (e^{2\phi_0}-1) V' (\phi_0) \nonumber \\
&&+ \frac{2\pi \epsilon G}{\epsilon +1} [\epsilon V' (\phi_0)
-2(2+\epsilon)] \int_0^{\phi_0} d\phi' \, e^{2\phi'}A_2(\phi')-
\epsilon [\widetilde{A}_3(\phi_0) -\widetilde{A}_3(0)], \label{8} \\
\beta_Z (\phi_0) &=& \mu {\partial Z (\phi_0)  \over \partial
\mu}
= 8\pi \epsilon A_1 G (e^{2\phi_0}-1) [Z' (\phi_0)-2]- \epsilon
[A(\phi_0)- A(0)]\nonumber \\
&&- \frac{2\pi \epsilon^2 G}{\epsilon +1} [e^{2\phi_0}A_2 (\phi_0)
-A_2(0)] + \frac{2\pi \epsilon^2 G}{\epsilon +1} [ Z' (\phi_0)
-1] \int_0^{\phi_0} d\phi' \, e^{2\phi'}A_2(\phi').
\label{9}
\end{eqnarray}
Here $\phi_0$ denotes the bare (non-renormalized) dilaton. We calculate
the dilatonic $\beta$-functions at $\phi_0$ because we are mainly
interested in the functional dilatonic dependence of these
$\beta$-functions. For consistency, in expressions (\ref{6})--(\ref{9})
terms with $\widetilde{A}_3(0)$ and $A_2(0)$ appear. As has been mentioned in
\cite{4,5} these terms are actually connected with the renormalization
of the zero modes of the dilatonic functions. The term with
$\widetilde{A}_3(0)$ in (\ref{8}) appears as a result of the renormalization of
the (non-essential for us) coupling constant $m^2$ \cite{5}. At the same
time, the terms that would carry $A(0)$ and $A_2(0)$ in (\ref{9})
disappear through renormalization of the trace of the metric, where they
are of next-to-leading order. (Recall that the renormalization of the
trace of the gravitational field at leading order is, according to
(\ref{5}), an ${\cal O} (1)$ term.)

Let us now search for fixed points of the beta functions. The
gravitational coupling constant shows asymptotic freedom, with the
corresponding ultraviolet fixed point being
\begin{equation}
G^* = \frac{3\epsilon}{2(24-n)}, \label{10}
\end{equation}
where $n<24$ and $\epsilon >0$. Hence, asymptotic freedom for $G$ is
obtained only when the matter central charge satisfies $0<n<24$
and $\epsilon \rightarrow 0$.
When searching for the fixed points of
the dilatonic couplings through the following {\it Ansatz}
\begin{equation}
\Phi (\phi) = \lambda \phi, \ \ \ \
V (\phi) = \lambda_V \phi, \ \ \ \
Z (\phi) = \lambda_Z \phi,
\label{11}
\end{equation}
we can write the beta functions under the form:
\begin{eqnarray}
\beta_\Phi &=& G(e^{2\phi} -1) \left[ \frac{24-n}{3} \lambda +
\frac{\epsilon}{4(1+\epsilon)} (\lambda -1) (n\lambda^2-8)
\right], \nonumber \\
\beta_V &=& G(e^{2\phi} -1) \left[ \frac{24-n}{3} \lambda_V +
\frac{n\lambda^2-8}{4(1+\epsilon)} (\epsilon\lambda_V-4 -
2\epsilon) -4\lambda_V -8 \right], \nonumber \\
\beta_Z &=& G(e^{2\phi} -1) \left[ \frac{12-n}{3} (\lambda_Z -2) +
\frac{\epsilon (n\lambda^2-8)}{4(1+\epsilon)} (\lambda_Z-3)
 \right].
\label{12}
\end{eqnarray}
A detailed study of these equations (\ref{12}) yields the following
results.
For the solution of Ref. \cite{4}
\begin{equation}
\lambda^* = -\frac{6\epsilon}{24-n} + {\cal O} (\epsilon^2),
\label{13}
\end{equation}
one has
\begin{equation}
\lambda^*_V = \frac{12\epsilon}{12-n} + {\cal O} (\epsilon^2), \ \ \ \
\ \ \ \  \lambda^*_Z = 2 + {\cal O} (\epsilon).
\label{14}
\end{equation}
Hence, the fixed point for $Z(\phi )$ appears at order ${\cal O} (1)$,
and it is not influenced by the explicit form of (\ref{13}).
As a whole, the system has a non-trivial fixed point in the
space of couplings, namely $(G^*,\lambda^* \phi, \lambda^*_V \phi,
\lambda^*_Z \phi)$, which turns out to be a saddle point. In fact,
when studying the
stability of the fixed point (\ref{10}), (\ref{13}), (\ref{14}),
we can perform
variations along four different trajectories.
A careful analysis of the
beta-functions (\ref{12}) for the linear Ansatz (\ref{11}) shows
that the two last equations (i.e. those for $V$ and $Z$) do not
produce a new multiplicity of solutions. In other words, for each
value of $G^*$ and $\lambda^*$ we just have one single value of
$\lambda_V^*$ and one of $\lambda^*_Z$, that complete the four
coordinates of the fixed point. For $\lambda^*$ we obtain three
distinct solutions: the real one (\ref{13}) and two purely
imaginary ones, of order $\epsilon^{-1/2}$, namely
\begin{equation}
\lambda^*_\pm = \pm 2i\sqrt{\frac{24-n}{3n\epsilon}} + {\cal O}
(\epsilon^0), \label{15}
\end{equation}
which correspond to highly oscillating dilaton couplings.
Then
\begin{equation}
\lambda^*_V = \frac{24-n}{12\epsilon} + {\cal O}
(\epsilon^0), \ \ \ \ \ \ \ \
\lambda^*_Z = \frac{48-n}{12} + {\cal O}
(\epsilon).
 \label{15p}
\end{equation}
When $\epsilon \rightarrow 0$ all dilatonic couplings are divergent,
except for $\lambda_Z^*$, which is finite as in the previous case. We
see that the explicit fixed point for the dilatonic coupling
$\lambda_Z^*$ appears always at order ${\cal O} (\epsilon^0)$, what is
quite a distinguished behavior from that of $\lambda^*_V$ and
$\lambda^*$. The fixed point (\ref{15}) is difficult to interprete as a
physically acceptable solution. Perhaps it indicates the possibility of
some kind of signature-changing transition (a transition to a dilatonic
model with a different signature). It would be also interesting to study
simulations of such a model (for an introduction, see \cite{28}) near
its fixed point.

Expanding the beta functions near the (only real) fixed point, in
the way
\begin{equation}
G=G^*+ \delta G, \ \ \ \ \Phi=\lambda^*\phi+ \delta \Phi, \ \ \ \
V=\lambda_V^* \phi+ \delta V, \ \ \ \ Z=\lambda^*_Z \phi +
\delta Z, \label{15pp}
\end{equation}
and assuming all the fluctuations to be small, we obtain
\begin{eqnarray}
\delta \beta_G &=&- \epsilon \delta G, \nonumber \\
\delta \beta_\Phi &=&\frac{ \epsilon}{2} \left(e^{2\phi} -
1\right) \frac{d}{d\phi} \delta \Phi + {\cal O} (\epsilon^2), \nonumber
\\
\delta \beta_{V}&=&\frac{ \epsilon}{2} \left( \frac{12-n}{24-n}
e^{2\phi} -1\right) \frac{d}{d\phi} \delta V + {\cal O}
(\epsilon^2), \nonumber \\
\delta \beta_Z&=&\frac{1}{2} \left(
e^{2\phi} -1\right) \frac{d}{d\phi} \delta Z + {\cal O}
(\epsilon). \label{263}
\end{eqnarray}
As observed in ref. \cite{4}, we may take $e^\phi$ to play the
role of loop expansion parameter, and restrict ourselves to the
region $e^{2\phi} \leq 1$, that is $-\infty < \phi \leq 0$. The
change of variables
\begin{equation}
\eta_1 =\ln (e^{-2\phi} -1), \ \ \ \ \ \
\eta_2 =\ln \left[ \frac{30-n}{12} \, \left( e^{-2\phi} -
\frac{18-n}{30-n}\right) \right], \label{264}
\end{equation}
transform this region into the following ones
\begin{eqnarray}
\phi =0, & \eta_1 \rightarrow -\infty,  & \eta_2 =0,  \nonumber \\
\phi \rightarrow -\infty, & \eta_1 \rightarrow +\infty,  & \eta_2
\rightarrow +\infty,  \label{265}
\end{eqnarray}
respectively. They simplify expressions (\ref{263}), which now
read:
\begin{eqnarray}
\delta \beta_G &=&- \epsilon \delta G, \nonumber \\
\delta \beta_\Phi &=& \epsilon  \frac{d}{d\eta_1} \delta \Phi +
{\cal O} (\epsilon^2), \nonumber
\\
\delta \beta_V&=& \epsilon  \frac{d}{d\eta_2} \delta V +
{\cal O} (\epsilon^2), \nonumber
\\
\delta \beta_Z &=&   \frac{d}{d\eta_1} \delta Z + {\cal
O} (\epsilon).
 \label{266}
\end{eqnarray}
A similar analysis to the one carried out in \cite{5}
shows that
the fixed points for $V$ and $Z$ are ultraviolet stable in
the direction $\delta \Phi$ but are always infrared unstable in this
direction. The point
 (\ref{10}), (\ref{13}), (\ref{14}),
 is a saddle fixed point of the RG equations.

Thus we have showed that the general dilatonic gravity theory (\ref{1})
has an asymptotically free regime for the gravitational coupling
constant
in which the theory possesses a non-trivial ultraviolet fixed point for
all dilatonic couplings.

We shall now see explicitly how the theory (\ref{1}) can be represented
in a very simple form at the non-trivial fixed point (\ref{13}),
(\ref{14}), displaying an asymptotically free gravitational coupling
constant. To this end let us perform a Weyl transformation of the action
as follows
\begin{equation}
g_{\mu\nu} \longrightarrow g_{\mu \nu} \exp \left(
\frac{4\lambda^*}{\epsilon} \phi \right).
\label{16}
\end{equation}
Then
\begin{eqnarray}
S &=&  \int d^d x\, \sqrt{-g}\,
\left\{ \frac{\mu^\epsilon}{16\pi G^*} e^{-2(1-\lambda^*)\phi}
\left[ R - \frac{4(1+\epsilon )}{\epsilon} \lambda^* (2-\lambda^*)
g^{\mu\nu} \partial_\mu \phi \partial_\nu \phi \right]
 - \frac{1}{2}
g^{\mu\nu} \partial_\mu \chi_i \partial_\nu \chi^i \right. \nonumber \\
&& + \left.
\mu^\epsilon
m^2 \exp\left[\left(2\lambda^* + \frac{4\lambda^*}{\epsilon} -
\lambda^*_V\right)\phi\right] + \frac{\mu^\epsilon}{2} \exp\left[ \left(
2\lambda^* -2 \lambda_Z \right) \phi \right]
g^{\mu\nu} \partial_\mu \phi \partial_\nu \phi \right\}.
\label{17}
\end{eqnarray}
As we see here, the general dilatonic gravity theory we have been
considering can be cast, at the non-trivial fixed point and near two
dimensions, under the form (\ref{17}), where the $n$ scalars do not
interact explicitly with the dilaton. However, though being of the
form of a string-inspired dilatonic model, (\ref{17}) does not represent
the CGHS action \cite{11}. In this respect, it is interesting to notice
that the particular case of the model (\ref{1}) with $\exp [-2Z(\phi)]
=0$ has indeed this feature: it also possesses a non-trivial ultraviolet
fixed point at which it can be represented by the CGHS action (see
\cite{4,5}).

The model (\ref{17}) is finite for $\epsilon\rightarrow 0$ (in exactly
two dimensions) and one can now study different properties of it, as
two-dimensional black hole solutions (see \cite{13,14}), once we get to
two dimensions.

 \bigskip

\section{Fermion-dilatonic gravity near two dimensions.}

We can easily extend the above picture to other forms of matter. As an
example we shall here consider the quite interesting case of dilatonic
gravity interacting with $m$ Majorana fermions. The corresponding
Lagrangian is
 \begin{equation}
L = \frac{\mu^\epsilon}{2} e^{-2Z(\phi)} g^{\mu\nu}
\partial_\mu \phi\partial_\nu \phi + {\mu^\epsilon \over 16\pi
G} e^{-2\phi} R- \frac{i}{2}  e^{-q(\phi)}
\bar{\psi}_a \gamma^\lambda \partial_\lambda \psi_a,
 \label{c1}
\end{equation}
where $\psi_a$ is an $m$-component Majorana spinor. The one-loop
renormalization of this theory can be done in close analogy with the
case of dilaton-scalar gravity. Using the one-loop counterterms found in
Ref. \cite{1}, we can write
\begin{equation}
\Gamma_{count} =- \mu^\epsilon  \int d^d x\,  \sqrt{-g}\,
\left[
RA_1 + g^{\mu\nu} \partial_\mu \phi \partial_\nu \phi
\bar{A}_2(\phi) + e^{-q(\phi )} \mu^{-\epsilon} A_3(\phi) T \right],
\label{c2}
\end{equation}
where
\begin{eqnarray}
&& T= -\frac{i}{2} \bar{\psi}_a \gamma^\lambda \partial_\lambda \psi_a,
\ \ \ \ \ \
A_1 = \frac{48-m}{48\pi\epsilon}, \nonumber \\
&&\bar{A}_2(\phi) = A_2 +e^{-2Z(\phi )} A(\phi ) =
-\frac{2}{\pi\epsilon} - \frac{4G}{\epsilon}
  e^{2\phi - 2Z(\phi)} [2 - Z'(\phi)], \nonumber \\
&&A_3(\phi) = -\frac{4G}{\epsilon} e^{2\phi} \left[\frac{3}{4}- \frac{
q'(\phi)}{2} +12\pi G e^{2\phi -2Z(\phi)} \right]. \label{c3}
\end{eqnarray}
The beta-functions have the same form as in (\ref{7})-(\ref{10})
(with $A_2 (\phi) =A_2$), and instead of $\beta_\phi$ and $\beta_V$ we
have just one beta function.
\begin{eqnarray}
\beta_q (\phi_0) &=& \mu {\partial q (\phi_0)  \over
\partial \mu} = 8\pi \epsilon A_1 G (e^{2\phi_0}-1) q'
(\phi_0) \nonumber \\
&&+ \frac{\pi \epsilon G}{\epsilon +1} A_2 (e^{2\phi_0}-1)
[\epsilon q' (\phi_0) -2 (\epsilon +1)]
-\epsilon [A_3(\phi_0) -A_3(0)]. \label{c4}
\end{eqnarray}
Now, using the linear Ansatz
\begin{equation}
Z(\phi )= \lambda_Z \phi, \ \ \ \ \ \
q(\phi )= \lambda \phi,
\label{c5}
\end{equation}
we get the ultraviolet fixed point
\begin{eqnarray}
G^*&=&\frac{3\epsilon}{48-m}, \ \epsilon >0,  \ m < 48, \nonumber \\
\lambda_Z &=& 2 + {\cal O} (\epsilon ), \ m \neq 24 \ (\lambda_Z =3,
\mbox{for} \ m=24), \nonumber  \\
\lambda_q &=& \frac{336}{48 +m} + {\cal O} (\epsilon ).
 \label{c6}
\end{eqnarray}
Hence, we obtain once more a non-trivial fixed point, but it is to be
noticed that the behavior of the fermion-dilaton coupling is
qualitatively different from the one corresponding to the scalar-dilaton
coupling (where $\lambda \equiv \epsilon + {\cal O} (\epsilon^2)$). In a
similar way, a more complicated theory of dilatonic gravity with both
scalar and fermionic matter could be considered.
 \bigskip

\section{Jackiw-Teitelboim and lineal gravities near two
dimensions.}
In this section we consider the one-loop renormalization of two popular
models of dilatonic gravity, namely the Jackiw-Teitelboim model \cite{8}
and lineal gravity \cite{9}, near two dimensions. We will study the
situation when such models of dilatonic gravity are connected with
scalar matter in the same way as in our starting Lagrangian (\ref{1}).
The corresponding Lagrangians are now:
\begin{equation}
L_{JT}= \frac{\mu^\epsilon}{16 \pi G} e^{-2\phi} \left(
R + \Lambda \right) -
 {1 \over 2} e^{-2 \Phi (\phi)} g^{\mu\nu} \partial_\mu \chi_i
\partial_\nu \chi^i
\label{a1}
\end{equation}
and
\begin{equation}
L_{lg}= \frac{\mu^\epsilon}{16 \pi G} \left( e^{-2\phi}
R + \Lambda \right) -
 {1 \over 2} e^{-2 \Phi (\phi)} g^{\mu\nu} \partial_\mu \chi_i
\partial_\nu \chi^i.
\label{a2}
\end{equation}
As one can infer from the discussion in the previous section, these two
models are not fixed points of the RG in $2+\epsilon$ dimensions. It is
therefore natural to consider the one-loop renormalization of these
models from the very beginning, in order to see how their specific
properties actually influence the renormalization structure. In
particular, one might expect that the renormalization of both models
should be performed in a quite different way ---as compared with the
renormalization of the general dilatonic model previously considered.

The calculation of the one-loop effective action is done in the same
gauge (\ref{2}) as before, with the following result:
\begin{equation}
\Gamma_{div}^{JT} = \frac{1}{4\pi \epsilon}  \int d^d x\,
\sqrt{-g}\,
\left\{ \frac{24-n}{6} R - \left[ 8-n \Phi'(\phi)^2
\right] g^{\mu\nu} \partial_\mu \phi \partial_\nu \phi + 4\Lambda
\right\} \label{a3}
\end{equation}
and
\begin{equation}
\Gamma_{div}^{lg} = \frac{1}{4\pi \epsilon}  \int d^d x\,
\sqrt{-g}\,
\left\{ \frac{24-n}{6} R - \left[ 8-n \Phi'(\phi)^2
\right] g^{\mu\nu} \partial_\mu \phi \partial_\nu \phi +
2\Lambda e^{2\phi} \right\}, \label{a4}
\end{equation}
respectively.
At first look it would seem that the two theories, (\ref{a3}) and
(\ref{a4}), are both non-renormalizable in $2+\epsilon$ dimensions.
However, the situation is not so simple as it looks. Indeed, to begin
with one can always consider a subclass of the theories under discussion
with some specific choices for the dilatonic function $\Phi (\phi )$,
and things could depend on this choice. On the other hand, the one-loop
effective action is, generally speaking, a gauge dependent quantity (for
an introduction to the effective action formalism in QG, see \cite{10}).

To illustrate the situation, let us first consider lineal gravity and
impose that $e^{-2\Phi (\phi )} =1$. In this case, in the dilatonic
kinetic term in (\ref{a4}) the $\phi$-dependence disappears completely.
Moreover, one can also choose a gauge of Landau type with gauge
parameter $\alpha =0$. Then, as it was shown in the last of Refs.
\cite{6}, the
second and the third terms in (\ref{a4}) disappear in this gauge (the
first term in (\ref{a4}) is a gauge independent quantity). As a result,
we obtain that lineal gravity (\ref{a2}) with scalar matter not
interacting with the dilaton explicitly is one-loop multiplicatively
renormalizable in a gauge of Landau type in exactly-two dimensions.
If $n=24$ then, in addition, it is one-loop finite
in $2+\epsilon$ dimensions.

For the Jackiw-Teitelboim model the situation is somewhat different.
With the same choice for $\Phi (\phi )$ as above, we do not know of any
gauge ---as the one found in Ref. \cite{6}--- in which both the second
and third terms in $\Gamma_{div}^{JT}$ disappear simultaneously (what
should not mean that it does not exist). In the gauge under
consideration one can choose $\Phi (\phi) = \sqrt{8/n} \, \phi$ in order
to remove the dilatonic kinetic term. However, the last term in
(\ref{a3}) is still present. Hence, it is very likely that, in fact, the
Jackiw-Teitelboim model is not one-loop renormalizable, either
in exactly-two or in $2+\epsilon$ dimensions.

An important question in relation with two-dimensional dilatonic gravity
is the existence of solutions of black hole type (for a review see
\cite{13,14}).  It would be of interest to understand something about
the structure of black holes in $2+\epsilon$ dimensions. In particular,
let us consider the Jackiw-Teitelboim model, where it is known that
there exists a regular, asymptotically flat black hole spacetime
\cite{24}. Starting from the action (\ref{a1}) (without the scalars
$\chi_i$, for simplicity), we expect that the regular black holes near
two dimensions will be described by a metric of Schwarzschild type with
the dilaton, of the form
\begin{eqnarray}
ds^2 &=& -\left( \frac{\Lambda}{2} r^2 -M \right) dt^2 +
\frac{dr^2}{\frac{\Lambda}{2} r^2 -M} + \left[ f(r) d\theta
\right]^2 \epsilon, \nonumber \\
\phi &=& r + {\cal O} (\epsilon ),
\label{22}
\end{eqnarray}
where $f(r)$ is some regular function of the radius. Of course, a
rigorous mathematical description of the classical theory for
non-integer $\epsilon$ is lacking.

Another issue connected with the two models above is the possibility of
a gauge formulation of the same, using an extended Poincar\'e group
theory (see for example \cite{9}). It would be again of interest to
study such questions in the ($2+\epsilon$)-dimensional formalism.

\bigskip

\section{Discussion.}
In this paper we have studied a general theory of dilatonic gravity with
$n$ scalars,
near two dimensions, at the quantum level.
The one-loop $\beta$-functions have been calculated and a non-trivial
ultraviolet fixed point of the theory has been found. At the fixed
point
the gravitational coupling constant of the theory is asymptotically
free, and the theory can be given a form in which the scalars do not
interact with the dilaton. By performing convenient transformations, one
can also give the action
some other, different forms at the fixed point. For instance,
the transformation
\begin{equation}
g_{\mu\nu} \longrightarrow g_{\mu \nu} \exp \left(
\frac{2\lambda_V}{2+\epsilon} \phi \right),
\label{d1}
\end{equation}
converts the theory (\ref{1}) with the dilatonic potential
$m^2e^{-\lambda^*_V \phi}$ into a theory with cosmological constant
$m^2$ (no dilatonic potential).
Fermion-dilatonic gravity near two dimensions has been considered too.

A further remark is connected with the possibility to increase the
number of scalars (maintaining always the regime of asymptotic freedom)
by adding to the theory a Yang-Mills action with a dilatonic coupling
\cite{5}
\begin{equation}
L_{YM} = \frac{1}{4} e^{-f_2 (\phi )} (G^a_{\mu\nu})^2, \ \  \ \ \ \
a=1,2, \ldots, N.
\end{equation}
Then, only $A_1$ is going to change in Eqs. (\ref{5}) and, as a result,
one can show  that the following ultraviolet stable fixed point appears
\begin{eqnarray}
G^* = \frac{3\epsilon}{2(24+6N-n)}, & &
\lambda^* =- \frac{6\epsilon}{24+6N-n}, \nonumber \\
\lambda_V^* = \frac{12\epsilon}{12+6N-n}, & &
\lambda_f^* =- \frac{36\epsilon}{12+6N-n},
\label{d2}
\end{eqnarray}
where we have chosen $f_2(\phi) =\lambda_f \phi$, and
$\lambda_Z^*$ does not change.

Hence, by increasing the dimension $N$ of the gauge
group we may increase the number of scalars in the theory, while keeping
always the condition that we are in the situation with $G$
asymptotically free. The presence of the first term in (\ref{1}) does
not modify this conclusion.

As a final remark let us mention that the ($2+\epsilon$)-dimensional
formalism has been shown \cite{25} to be very useful in understanding
the gravitational dressing of the RG beta-function \cite{26} in the case
of  the sigma model interacting with two-dimensional Einstein gravity.
It
is a challenge to understand the gravitational dressing of the RG in the
sigma model with dilatonic gravity. The results of our discussion here
are expected to be quite helpful in this respect.

 \vspace{5mm}


\noindent{\large \bf Acknowledgments}

SDO  would like to thank R. Kantowski, S. Naftulin, N. Sakai, Y.
Tanii and R.W. Tucker for very interesting discussions.
This work has been supported by DGICYT (Spain), project Nos.
PB93-0035
and SAB93-0024,  by CIRIT (Generalitat de Catalunya), and by RFFR
(Russia), project No. 94-02-03234.

\end{document}